\documentclass[conference]{IEEEtran}
\IEEEoverridecommandlockouts
\usepackage{cite}
\usepackage{amsmath,amssymb,amsfonts}
\usepackage{algorithmic}
\usepackage{graphicx}
\usepackage{textcomp}
\usepackage{xcolor}
\usepackage{hyperref}

\newboolean{showcomments}
\setboolean{showcomments}{false}
\ifthenelse{\boolean{showcomments}}
{ \newcommand{\mynote}[3]{
     \fbox{\bfseries\sffamily\scriptsize#1}
        {\small$\blacktriangleright$\textsf{\emph{\color{#3}{#2}}}$\blacktriangleleft$}}}
{ \newcommand{\mynote}[3]{}}
\definecolor{asparagus}{rgb}{0.53, 0.66, 0.42}

\def\BibTeX{{\rm B\kern-.05em{\sc i\kern-.025em b}\kern-.08em
    T\kern-.1667em\lower.7ex\hbox{E}\kern-.125emX}}

\begin{document}


\title{The World Model Planning Accelerator (WMPA): Hardware-Algorithm Co-Design for Embedded AI in Robotics}

\author{
\IEEEauthorblockN{Prerana Ramkumar}
\IEEEauthorblockA{
\textit{College of Engineering,} \\
\textit{Department of Computer Engineering,} \\
\textit{American University of Sharjah} \\
Sharjah, United Arab Emirates \\
g00100339@aus.edu}
\and
\IEEEauthorblockN{Suhas Suresh Bharadwaj}
\IEEEauthorblockA{
\textit{Department of Electrical} \\
\textit{and Electronics Engineering,} \\
\textit{BITS Pilani -- Dubai Campus} \\
Dubai, United Arab Emirates \\
f20230029@dubai.bits-pilani.ac.in}
\and
\IEEEauthorblockN{Fazal Hameed}
\IEEEauthorblockA{
\textit{College of Engineering,} \\
\textit{Department of Computer Engineering,} \\
\textit{American University of Sharjah} \\
Sharjah, United Arab Emirates \\
fhameed@aus.edu}
}

\maketitle

\begin{abstract}
Embedded AI is moving from perception to prediction. Robots now use world models---learned simulators that predict the outcome of candidate actions in a compressed latent space before any is executed. Every predicted step must be computed on board within a tight power and thermal budget, yet the energy these steps fundamentally require, and how far real hardware sits from it, have not been characterized. We derive lower bounds on the arithmetic operations and the CMOS dynamic energy of a latent state transition from information theory and Boolean circuit complexity. Profiling a DreamerV3-style Recurrent State Space Model (RSSM) shows that every rollout operation is memory-bandwidth bound and that a GPU consumes at least $5.9\times10^{10}$ times the CMOS energy floor. Transient 3D thermal simulations show that this inefficiency is also a thermal one, since sustained back-to-back planning heats a millimeter-scale silicon tile to 556~K even behind an ideal copper heat spreader. Guided by this chain from bound to bottleneck, we propose the World Model Planning Accelerator (WMPA), a chip architecture for embedded world model planning that combines near-memory computation, real-time entropy estimation, variable-precision rollout units, and copy-on-write latent scratchpads. With a design target of $40\times$ lower power, the same tile stays near ambient at 306~K in simulation, pointing toward practical planning on thermally constrained robots.
\end{abstract}

\begin{IEEEkeywords}
World Models, Embedded AI, Robotics, Hardware Accelerator, Memory-Bound, Energy Efficiency
\end{IEEEkeywords}

\section{Introduction}

Embedded AI in robotics increasingly relies on world models to rehearse actions internally before committing to them, which is valuable when real-world trials are slow, costly, or unsafe. A world model is a parameterized function $f : \mathcal{Z} \times \mathcal{A} \rightarrow \mathcal{Z}$ that takes a latent state $z_t \in \mathcal{Z}$ and an action $a_t \in \mathcal{A}$ and returns a predicted successor $z_{t+1} \in \mathcal{Z}$. Planners such as Model Predictive Control (MPC) \cite{hansen2022temporal} and Monte Carlo Tree Search (MCTS) \cite{schrittwieser2020mastering} apply this step $H$ times across $B$ parallel trajectories and score the imagined futures to choose a physical action. 

Work on world models has been largely algorithmic, with DreamerV3 \cite{hafner2025mastering} showing that Recurrent State Space Models (RSSMs) work across domains without hyperparameter tuning. Hardware work has targeted static inference, from the Tensor Processing Unit (TPU) \cite{jouppi2017in}, whose systolic arrays amortize memory accesses over static matrix multiplications, to Processing-in-Memory (PIM) \cite{mutlu2019processing}, which relieves off-chip bandwidth.

Yet neither effort has characterized the hardware limits of latent rollout planning. Unlike the feed-forward inference of a Convolutional Neural Network (CNN) or a static Transformer, a rollout is a deep sequential chain in which every step waits on the one before it, so wide systolic arrays sit idle and memory bandwidth becomes the constraint. This matters most for the embedded processors of edge robots \cite{lin2018architectural}, where thermal envelopes and power budgets are strict, and it raises two questions. What is the least energy a rollout can dissipate, and how far above it does deployed hardware sit?

This paper makes the following contributions:
\begin{itemize}
    \item \textbf{(1) Theoretical Hardware Bounds.} Under the Markovian and structural assumptions stated in Section II, we use Boolean circuit complexity to lower bound the arithmetic operations per step by the information the step must carry, and convert that count into an energy floor through the dynamic switching energy of CMOS gates. Both bounds grow linearly in $H$ and $B$.
    \item \textbf{(2) Empirical Architectural Profiling.} We profile a DreamerV3-style RSSM with a roofline model and a sustained-load \texttt{pynvml} protocol. Every core rollout operation is memory-bandwidth-bound, and the measured energy sits about eleven orders of magnitude above the CMOS floor.
    \item \textbf{(3) Transient Thermal Constraints.} Under sustained back-to-back planning and an assumed 9\% hotspot power share, GPU-class power density drives a mm-scale silicon tile to an unsustainable $556.07\text{ K}$ even with an idealized copper heat spreader. 
    \item \textbf{(4) Accelerator Co-Design.} We propose the World Model Planning Accelerator (WMPA), a chip architecture whose entropy-driven precision scaling and near-memory compute aim at a $40\times$ power reduction that would hold the same tile at $306.52\text{ K}$.
\end{itemize}

Near-memory computing \cite{mutlu2019processing} and reduced-precision arithmetic \cite{micikevicius2018mixed} are individually well established, yet all target static CNN and Transformer inference, whereas a rollout's computation follows the mutual information that survives each latent transition. Contributions (1) to (3) therefore form one chain from a mutual-information bound to the measured energy gap and its thermal limit, and the accelerator (4) is designed around the bottleneck this chain exposes. The bounds of contribution (1) are not design targets but the lowest point any implementation could reach.

The profiling and roofline code is publicly available at https://github.com/Prk10/world-models-theoretical-min.

\section{Theoretical Background}

\subsection{Entropy Growth Under Open-Loop Rollout}
\label{subsec:EntGrowth}

\textbf{Assumption A1.} (Latent Markov property) The latent state $Z_t$ is a sufficient statistic for predicting $Z_{t+1}$ under a fixed action sequence, so an open-loop rollout without observations follows the joint distribution
\begin{equation}
    p(z_{1:H} | a_{1:H-1}) = p(z_0) \prod_{t=0}^{H-1} p_\theta(z_{t+1} | z_t, a_t) \label{eq:markov}
\end{equation}

\textbf{Assumption A2.} (Stochastic transition) The transition model $p_\theta(z_{t+1} | z_t, a_t)$ has strictly positive variance and bounded logits, which we take to give a categorical latent space $\mathcal{Z}$ a minimum per-step conditional entropy $\delta > 0$ satisfying

\begin{align}
    \mathcal{H}(Z_{t+1} \mid Z_t) \geq \delta \label{eq:conditional_entropy}
\end{align}


Under A1 the action sequence is treated as fixed and known, so $(Z_0, Z_t, Z_{t+1})$ forms a Markov chain and the data processing inequality \cite{cover2006elements} applies. This gives a guarantee weaker than unbounded entropy growth. Since an open-loop rollout never reintroduces observations, $I(Z_0; Z_{t+1}) \leq I(Z_0; Z_t)$, or equivalently $\mathcal{H}(Z_t \mid Z_0)$ is non-decreasing in $t$ until it saturates at the entropy ceiling of the latent alphabet.

\begin{align} 
    \mathcal{H}(Z_t \mid Z_0) \leq \mathcal{H}(Z_{t+1} \mid Z_0) \leq \log_2 |\mathcal{Z}| \quad \text{for all } t \geq 0 
    \label{eq:entropy_growth} 
\end{align}


For the $32 \times 32$ categorical latent studied here the ceiling is $\log_2 |\mathcal{Z}| = 160$ bits (Section \ref{Subsec:NumAnalysis}), the largest information the latent can carry, so the bounds of Sections \ref{Sec:TheAnalysis} and \ref{Sec:EmpProfiling} are evaluated at this value and not assumed to grow without limit. The step from a non-increasing $I(Z_0;Z_t)$ to a non-decreasing $\mathcal{H}(Z_t \mid Z_0)$ also needs $\mathcal{H}(Z_t)$ itself to be non-decreasing, which we take as a structural assumption for the RSSM studied here and do not prove.

\textit{Remark:} A1 defines the open-loop condition, since closed-loop planning with posterior updates reduces entropy and needs a separate analysis. A2 fails for deterministic transitions but holds by construction for the stochastic categorical latent of the RSSM studied here.

\subsection{The Roofline Model}
The roofline model \cite{williams2009roofline} tells whether a workload is limited by compute throughput or by memory bandwidth, through its arithmetic intensity
\begin{equation}
    \mathrm{AI} = \frac{\text{FLOPs}}{\text{Bytes transferred}} \label{eq:intensity}
\end{equation}


A device with peak throughput $\Pi$ (FLOPs/s) and peak bandwidth $\beta$ (bytes/s) has a ridge point $\rho = \Pi / \beta$ (FLOPs/byte). Below it ($\mathrm{AI} < \rho$) a workload is memory-bandwidth-bound and its performance follows $\beta$, so more compute throughput without more bandwidth gains nothing. Above it ($\mathrm{AI} > \rho$) the workload is compute-bound and limited by $\Pi$.

\subsection{Information Theory and Landauer's Principle}

Shannon entropy $\mathcal{H}(Z)$ measures the average information in a random variable $Z$ \cite{cover2006elements} and is given by $\mathcal{H}(Z) = -\sum_{z} p(z)\log_2 p(z)$. The mutual information $I(Z_t; Z_{t+1})$ between consecutive latent states then gives the information that any transition computation must carry. Both quantities depend on the learned model and not on the hardware.

Landauer's principle \cite{landauer1961irreversibility} ties computation to thermodynamics by requiring any logically irreversible operation, such as erasing a bit, to dissipate at least
\begin{equation}
    E_{\min} = k_B T \ln 2 \label{eq:landauer}
\end{equation}


\noindent as heat, where $k_B$ is Boltzmann's constant and $T$ is the operating temperature. At $T \approx 300$ K this is about $2.87 \times 10^{-21}$ J per bit erased, a thermodynamic floor that real solid-state devices exceed by a wide margin.
\section{Theoretical Analysis}
\label{Sec:TheAnalysis}

We lower bound the cost of a single rollout step through the mutual information $I(Z_t; Z_{t+1})$ between consecutive latent states, which is the information the transition computation must propagate. By Section \ref{subsec:EntGrowth}, every step adds at least $\delta$ bits of new conditional entropy.

\subsection{Arithmetic Complexity Bound on MAC Operations}\label{subsec:MACBound}

\textbf{Assumption A3.} \textit{(Information flow in ALUs)} The computational primitive of the transition $z_t \rightarrow z_{t+1}$ is the Multiply-Accumulate (MAC) operation of an Arithmetic Logic Unit (ALU), which processes $b$-bit operands, and every bit of information carried from $z_t$ to $z_{t+1}$ passes through at least one gate evaluation, so free wiring is not credited.

\textbf{Proposition 1.} \textit{(Gate-count floor)} Under A3, the number of logic gates that a circuit implementing the transition $z_t \rightarrow z_{t+1}$ evaluates per step satisfies

\begin{equation}
    N_{gates}^{\min} \geq I(Z_t; Z_{t+1}) \label{eq:gate_bound}
\end{equation}

At the MAC level, each $b$-bit operand carries at most $b$ bits, so propagating $I(Z_t;Z_{t+1})$ bits through such a datapath takes at least

\begin{equation}
    \text{OPS}_{step}^{\min} \geq \frac{I(Z_t; Z_{t+1})}{b} \label{eq:mac_bound}
\end{equation}
For a planning call with $B$ trajectories over horizon $H$ this becomes
\begin{equation}
    \text{OPS}_{total}^{\min} \geq \frac{B \times H \times I(Z_t; Z_{t+1})}{b} \label{eq:mac_total_bound}
\end{equation}

A $b$-bit MAC is realized with $O(b^2)$ gates in standard array multipliers \cite{savage1998models}. We use this count, $K(b) \propto b^2$, only as a realization cost and not as a proven lower bound, since superlinear lower bounds for multiplication remain an open problem. It is applied only in Section \ref{subsec:CompDesign} to compare ALU precisions and does not enter Eqs. \eqref{eq:gate_bound}--\eqref{eq:mac_total_bound}.

\textit{Remark:} Eqs. \eqref{eq:mac_bound}--\eqref{eq:mac_total_bound} show the minimum MAC count growing as $1/b$, but each MAC costs $K(b) \propto b^2$ gates, so the total gate cost scales as $b \, I(Z_t;Z_{t+1})$ and falls with precision. Low precision need not be optimal for task performance, since quantization error may require extra rollout steps.

\subsection{Thermodynamic Lower Bound on Energy (Proposition 2)}
\label{subsec:ThermLower}

\textbf{Assumption A4.} (Logical irreversibility) Each step overwrites $z_t$ with $z_{t+1}$ in memory, so $z_t \rightarrow z_{t+1}$ is many-to-one over possible trajectories.

\textbf{Proposition 2.} (Thermodynamic energy bound) Under A4, and treating the sampling noise as an additional input of the update, overwriting $z_t$ with $z_{t+1}$ discards at least $\mathcal{H}(Z_t \mid Z_{t+1}) = \mathcal{H}(Z_t) - I(Z_t;Z_{t+1})$ bits, so by Landauer's principle a step dissipates at least $E_{step} \geq \mathcal{H}(Z_t \mid Z_{t+1}) \, k_B T \ln 2$. This quantity vanishes for an information-preserving update and does not follow from $I(Z_t;Z_{t+1})$ alone, so unlike Proposition 3 it is a model-dependent reference scale that we state without evaluating.

\subsection{Physical Limits of CMOS Energy Dissipation}

Landauer's limit is the ultimate floor, but it ignores the physics of semiconductor switching. A more realistic baseline is the minimum dynamic energy of the transistors themselves.

\textbf{Assumption A5.} (CMOS switching limit) Each gate transition on a CMOS substrate charges and discharges a parasitic capacitance $C_{\text{gate}}$ at supply voltage $V_{dd}$.

\textbf{Proposition 3.} (Dynamic energy floor) The energy of one rollout step in a standard CMOS process is at least the minimum gate evaluations $N_{gates}$ of Proposition 1 times the switching energy per gate
\begin{equation}
    E_{step}^{CMOS} \geq I(Z_t; Z_{t+1}) \cdot \left( \alpha C_{\text{gate}} V_{dd}^2 \right) \label{eq:cmos_limit}
\end{equation}
where $\alpha \in (0,1]$ is the switching activity \cite{rabaey2002digital}.

\subsection{Numerical Evaluation of the Bounds}
\label{Subsec:NumAnalysis}

For the RSSM profiled in Section \ref{Sec:EmpProfiling}, the $32 \times 32$ categorical latent has a maximum Shannon entropy of exactly $160$ bits ($32 \times \log_2 32$), so $I(Z_t; Z_{t+1}) \le 160$ bits per step. We evaluate the floor at this ceiling.

Substituting this into Proposition 3 with TSMC 12nm FinFET parameters \cite{tsmc12nm} ($V_{dd} = 0.8\text{ V}$, $C_{gate} \approx 1\text{ fF}$, $\alpha = 0.1$) gives
\begin{equation}
    E_{step}^{CMOS} \geq 160 \times (0.1 \times 10^{-15} \times 0.8^2) = 1.024 \times 10^{-14} \text{ J} \label{eq:cmos_eval}
\end{equation}
and for a planning call with $B = 32$ and $H = 20$, counting each of the $B \times H$ transitions as distinct
\begin{equation}
    E_{plan}^{CMOS} \geq 32 \times 20 \times 1.024 \times 10^{-14} \approx 6.55 \times 10^{-12} \text{ J} \label{eq:cmos_total_eval}
\end{equation}

Measured on a 12nm GPU at the same $B = 32$ and $H = 20$, a planning call consumes $3.873\times10^{-1}$ J (mean of 10 trials, $\sigma = 5.923\times10^{-3}$ J, Fig. \ref{fig3}). Since $I(Z_t;Z_{t+1}) \le 160$ bits, the floor above is the largest value the bound can take, so for these device parameters the measured energy is at least $5.91\times10^{10}$ times the CMOS floor, roughly eleven orders of magnitude. The Landauer scale of $2.87\times10^{-21}$ J per bit lies more than four orders below the $6.4\times10^{-17}$ J switching energy per gate used here, so the CMOS floor is the binding one.

\section{Empirical Profiling of General-Purpose Architectures}
\label{Sec:EmpProfiling}

Section \ref{Sec:TheAnalysis} puts the energy floor many orders of magnitude below that of practical systems. Here we profile an RSSM on a GPU to see where the gap comes from and whether the dissipated energy can leave a millimeter-scale tile.

\subsection{Experimental Setup}

The RSSM follows \cite{hafner2025mastering} and is implemented in PyTorch. It has a 512-unit GRU as its deterministic state, a $32 \times 32$ categorical stochastic latent that flattens to 1024 dimensions, and a linear head that maps both states to a 512-dimensional feature.

All experiments run on a dedicated, isolated NVIDIA Tesla T4 GPU with $B = 32$ and $H \in \{5, 10, 15, 20\}$. Byte traffic for the roofline analysis (Section \ref{subsec:GenRoof}) is computed analytically from weight and activation sizes, and energy is measured under sustained load with \texttt{pynvml} power monitoring, averaged over 10 independent trials. The profile targets the structure of the rollout and not learned weights.

\subsection{Generational Roofline Analysis of Rollout Operations}
\label{subsec:GenRoof}

We project the rollout operations onto the roofline of three NVIDIA generations, the T4, A100, and H100 \cite{nvidia2022h100}. Arithmetic intensity is the ratio of FLOPs to minimum DRAM traffic, with the softmax exponentiation counted as a 15-FLOP Special Function Unit (SFU) cost.

\begin{figure}[htbp]
\centerline{\fbox{\includegraphics[width=0.95\linewidth]{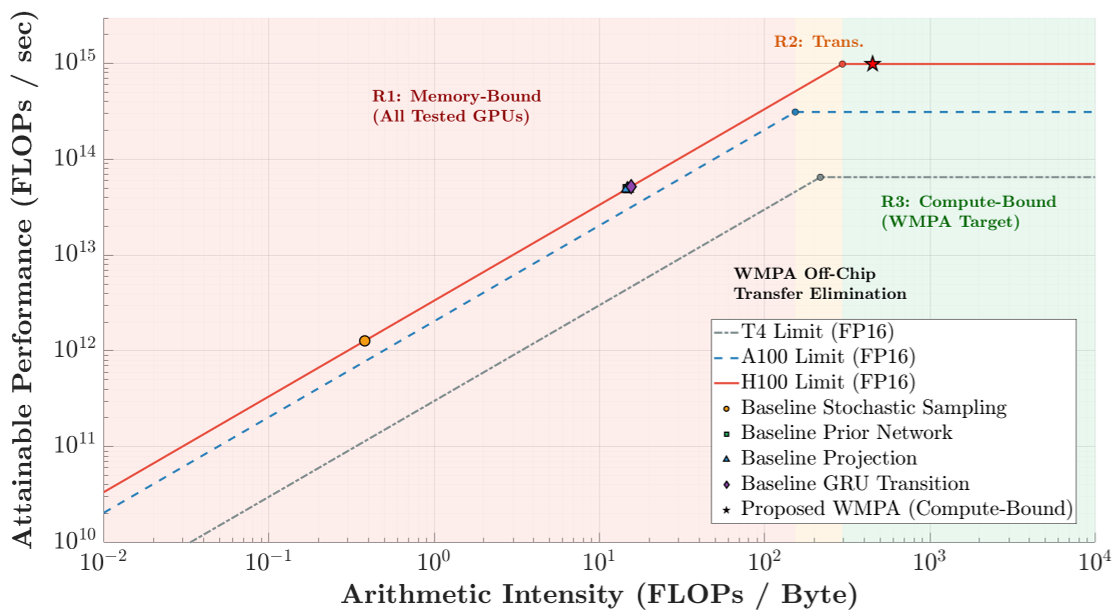}}}
\caption{Roofline projection of the rollout operations on three GPU generations (FP16 Tensor Core path). Ridge points are 203.5 (T4), 153.0 (A100), and 295.2 FLOPs/byte (H100), all for dense FP16. All operations (0.38 to 15.55 FLOPs/byte) lie in the memory-bound region R1 on every GPU.}
\label{fig1}
\end{figure}

As Fig. \ref{fig1} shows, the ridge points divide the plane into a memory-bound region R1, a transition region R2, and a compute-bound region R3. The arithmetic intensity of the workload lies between $0.38$ and $15.55\text{ FLOPs/byte}$, well below even the lowest ridge point tested (153.0 FLOPs/byte for the A100), so it stays in R1 on every GPU at every precision. 

Since no tested GPU leaves R1, extra FLOP/s cannot help, and the remedy is an architecture that raises effective arithmetic intensity and pushes the workload toward R3. This is the origin of the Near-Memory Compute Unit (NMCU) and Latent State Scratchpad (LSS) in Section \ref{Sec:PropAccelerator}.

\subsection{Dynamic Activation Sparsity and Sustained Energy Profiling}
\label{subsec:DynActSparsity}

Activation sparsity of a trained policy depends on the downstream environment, so we use an analytic stress test as an environment-agnostic baseline. A converged world model commits to a trajectory, so its per-step predictive entropy $\mathcal{H}(Z_{t+1} \mid Z_t)$ falls toward $\delta$ while the cumulative $\mathcal{H}(Z_t \mid Z_0)$ of Section \ref{subsec:EntGrowth} keeps growing. We emulate this by progressively decaying the softmax temperature \cite{hinton2015distilling} over an $H = 20$ rollout. This yields an illustrative sparsity trajectory under strong entropy collapse and not a bound on what a trained model would show.

\begin{figure}[htbp]
\centerline{\fbox{\includegraphics[width=0.95\linewidth]{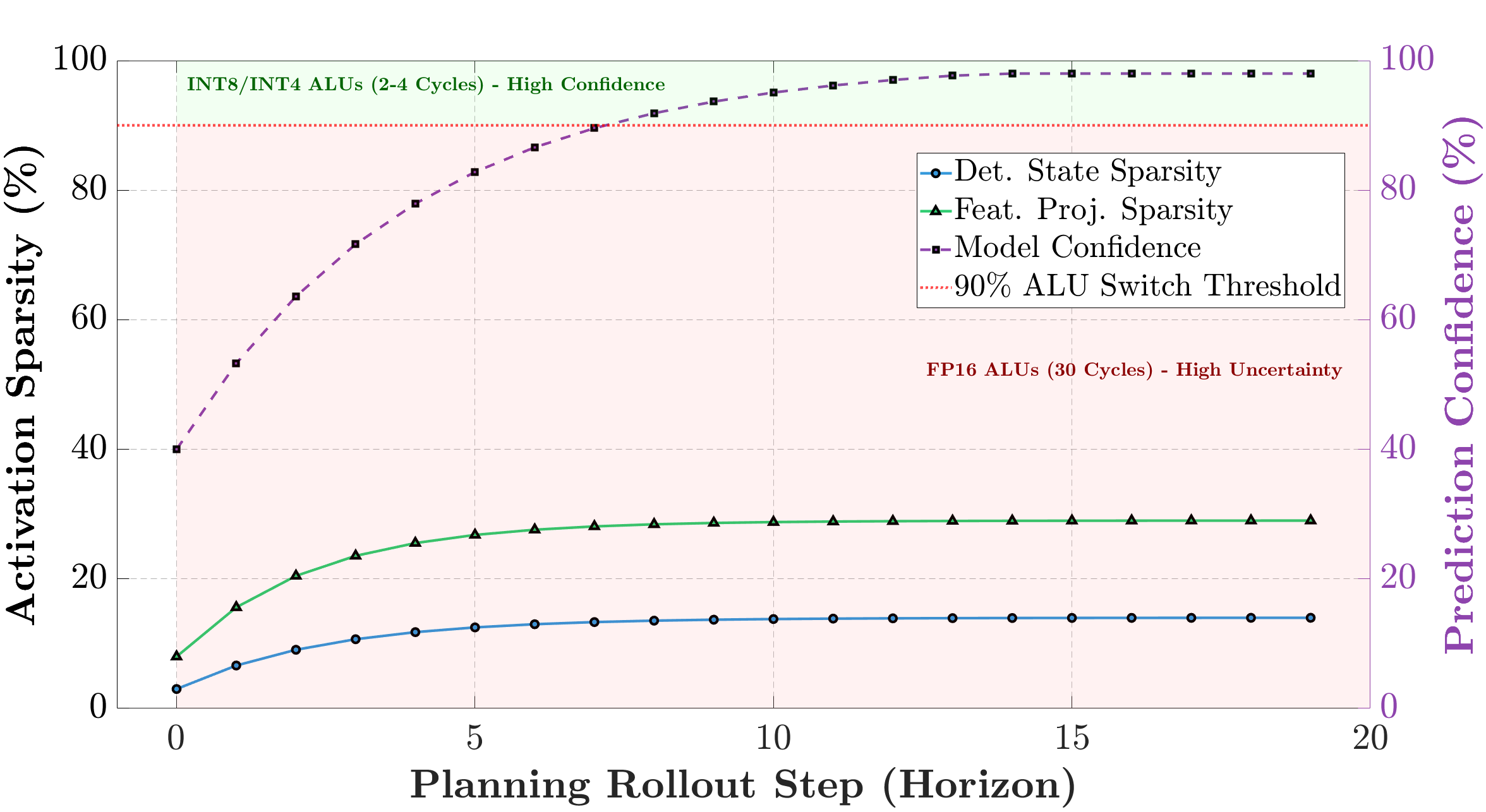}}}
\caption{Dynamic activation sparsity (values $< 0.01$) under the analytic softmax-temperature-decay stress test, a synthetic proxy and not a trained-policy measurement. Sparsity stabilizes at 14.0\% for the deterministic state and 29.0\% for the feature projection.}
\label{fig2}
\end{figure}

As Fig. \ref{fig2} shows, sparsity climbs as entropy drops over the horizon and settles at 14.0\% for the deterministic GRU state and 29.0\% for the feature projection. GPUs execute dense tensor operations, so near-zero elements still pay the full memory-access and MAC cost. This is the regime in which reduced precision becomes attractive, and it motivates the dynamic precision scaling of the Variable Precision Rollout Units (VPRU).

\noindent\textbf{Sustained Power and Energy Telemetry.} Multiplying execution time by the rated TDP misstates the energy of a memory-bound workload whose compute cores often stall. We therefore read board-level power-sense telemetry through \texttt{pynvml} every 50 ms across a continuous 3-second robotic planning loop.
The measurements are compared with the dynamic switching floor $E_{step}^{CMOS}$ of Proposition 3 for the TSMC 12nm FFN process of the T4 die, not with an idealized thermodynamic limit.

\begin{figure}[htbp]
\centerline{\fbox{\includegraphics[width=0.95\linewidth]{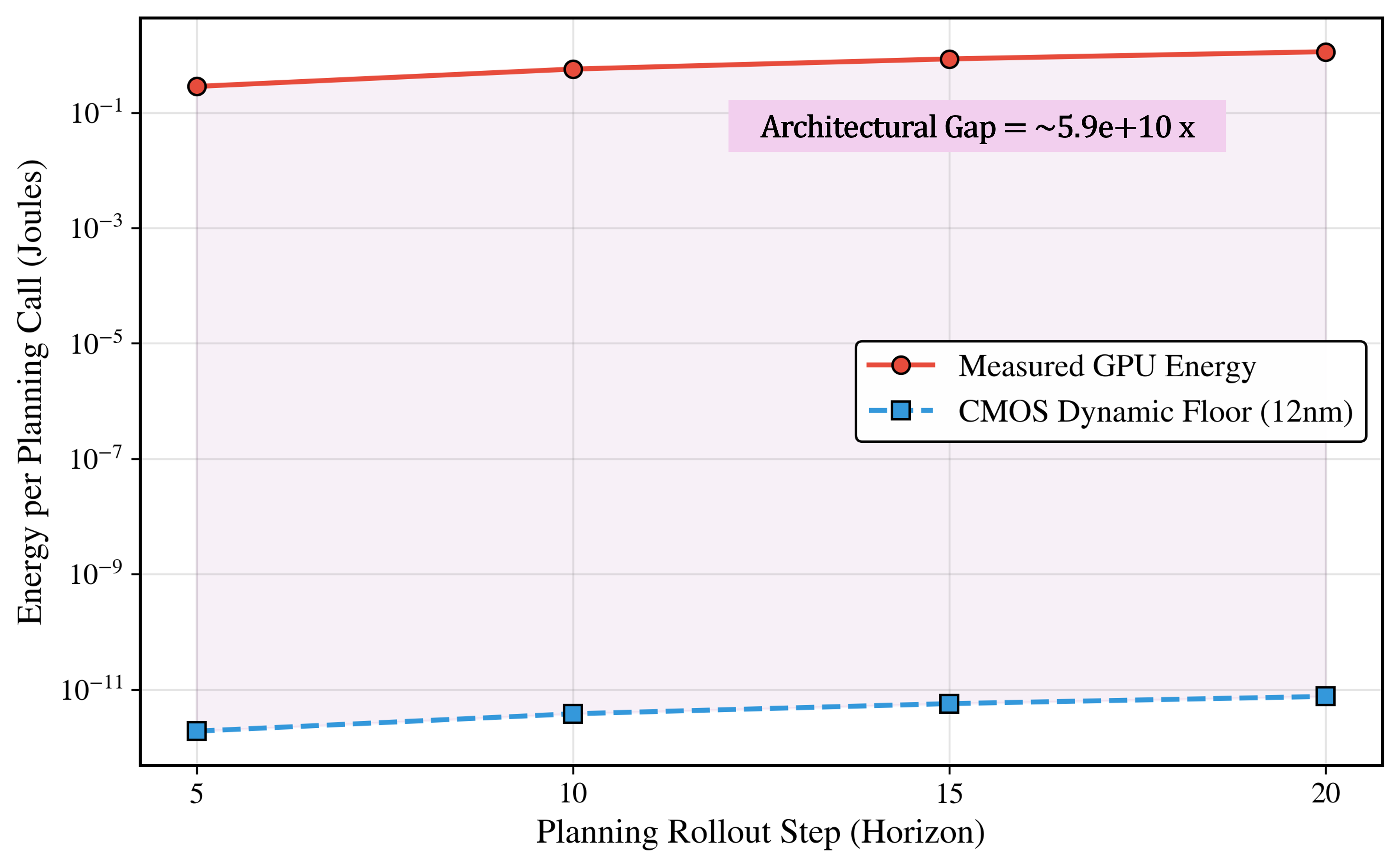}}}
\caption{Measured sustained GPU energy per planning call (mean of 10 trials, with standard deviation) against the 12nm CMOS dynamic switching floor. The gap is $5.91\times10^{10}$ at $H=20$, the headline configuration of Section \ref{Subsec:NumAnalysis}, and rises to $6.08\times10^{10}$ at shorter horizons because of a fixed per-call overhead.}
\label{fig3}
\end{figure}

Fig. \ref{fig3} shows three findings. First, the board draws 68.19 W, close to the 70 W rated TDP of the T4, even though the roofline analysis leaves the compute units largely idle. Second, energy scales linearly with $H$, as the bounds of Section \ref{Sec:TheAnalysis} predict. Third, the gap to the 12nm CMOS floor is $5.91\times10^{10}$ at $H=20$. Since $E_{step}^{CMOS}$ bounds only data-path transistors, while the 68.19 W board power also covers GDDR6 DRAM, voltage regulation, and interconnect, the gap reflects the whole subsystem and not a clean logic-versus-logic comparison, although the low compute utilization of Section \ref{subsec:GenRoof} points to data movement as the main contributor. This motivates moving computation next to the memory arrays through the NMCU.

\subsection{Transient Non-Linear Thermal Dynamics}
\label{subsec:TransientNonLinear}

The dissipated energy appears as localized Joule heating. To test whether such a compute density is thermally feasible in edge robotics, we solve the 3D transient non-linear heat conduction equation

\begin{equation}
    \rho_m C_p \frac{\partial T(\mathbf{r},t)}{\partial t} = \nabla \cdot \Big( k(T) \nabla T(\mathbf{r},t) \Big) + q_v(\mathbf{r},t) \label{eq:heat_conduction}
\end{equation}

where $\rho_m$ is the material density, $C_p$ the specific heat capacity, $T(\mathbf{r},t)$ the temperature field, and $q_v(\mathbf{r},t)$ the localized volumetric heat generation.

The thermal conductivity of silicon, $k(T)$, degrades at elevated temperature. Following the measurements of Glassbrenner and Slack \cite{glassbrenner1964thermal}, we model it empirically as

\begin{equation}
    k(T) = k_{300} \left( \frac{300}{T} \right)^\gamma \label{eq:thermal_cond}
\end{equation}

where $k_{300} \approx 152 \text{ W/m}\cdot\text{K}$ and $\gamma \approx 1.35$ come from a log-log regression of their data between $300$ and $600$ K ($R^2 = 0.994$).
 
For the worst case we use the longest tested horizon, $H = 20$, and sustained back-to-back planning calls. At $0.387$ J per call this corresponds to about 176 calls per second, well above the 30 Hz of typical real-time robotic control, where the planning energy alone would average only about 11.6 W. The baseline scenario is therefore a throughput-bound worst case and not a 30 Hz operating point. 

The baseline flux $Q_{baseline} = 6.17 \times 10^6 \text{ W/m}^2$ is not measured directly. It follows from assigning about 9\% of the measured 68.19 W board power to a $1 \text{ mm} \times 1 \text{ mm}$ near-memory compute tile. Since Eq. \eqref{eq:heat_conduction} needs a volumetric source, we place this dissipation in the active volume of the tile with a $t_{Si} = 100 \mu\text{m}$ silicon layer, so $q_v = Q / t_{Si}$. Its top surface exchanges heat with a 300 K ambient through an effective conductance $h_{eff} = 23{,}850 \text{ W/(m}^2\cdot\text{K)}$, the series resistance of a $1.5\text{ mm}$ copper spreader ($k_{Cu} \approx 385 \text{ W/(m}\cdot\text{K)}$) and a thermal interface material of $0.38 \text{ K}\cdot\text{cm}^2\text{/W}$, which idealizes the heat sink behind the spreader as held at ambient.

For the baseline GPU-class load (Scenario A) we treat the tile as a concentrated hotspot inside a streaming multiprocessor, since no floorplan-resolved power map of the T4 is available. The 9\% share is an order-of-magnitude assumption, and since the temperature rise is linear in $Q$ the 556.07 K peak scales with it, so we read it as a stress case and not a measured temperature. It gives $q_{v, baseline} = 6.17 \times 10^{10} \text{ W/m}^3$. For the WMPA (Scenario B), the design-target $40\times$ reduction gives $q_{v, WMPA} = 1.54 \times 10^9 \text{ W/m}^3$. Both sources are applied to the FEM model and evaluated at $t = 5.0\text{ s}$, which is far longer than the thermal time constant of the modeled tile, so the reported peaks are effectively steady-state values.

Scenario A rises to a peak of $556.07 \text{K}$, which is $256.07\text{ K}$ above ambient, whereas Scenario B stays at $306.52\text{ K}$, a rise of only $6.52\text{ K}$. As a consistency check, the steady-state one-dimensional balance $\Delta T \approx Q/h_{eff}$ gives $258.7\text{ K}$ and $6.5\text{ K}$ for the two scenarios, within about 1\% of the FEM rises. The rise is therefore set mainly by the heat-spreading boundary, with the degradation of $k(T)$ a secondary effect.

A sustained $\sim 550\text{ K}$ raises subthreshold leakage and, following Black's model \cite{black1969electromigration}, shortens electromigration lifetime, a serious risk of hardware failure. Even with this idealized spreader the baseline power density cannot be sustained, whereas the WMPA target stays near ambient. These results assume an ideal sink behind the spreader and a hotspot flux estimated from a flat 9\% power share, so they establish relative feasibility and not absolute operating temperatures.

\section{Proposed Accelerator Architecture}
\label{Sec:PropAccelerator}
\subsection{Design Principles and Overview}
\label{subsec:DesignPrinciples}

Five observations from the preceding sections, spanning theoretical bounds, hardware measurements, and an analytic proxy, motivate the WMPA. First, all rollout operations have $\mathrm{AI} < \rho$ on all three GPUs (Section \ref{subsec:GenRoof}), so extra FLOP/s do not improve throughput. Second, in the analytic stress test of Section \ref{subsec:DynActSparsity}, a synthetic proxy and not a trained-policy measurement, the GRU state reaches $14.0\%$ sparsity and the feature projection $29.0\%$, none of which dense hardware exploits. Third, the CMOS floor of Proposition 3 grows linearly with $I(Z_t; Z_{t+1}) \le \mathcal{H}(Z_t)$, which sets the minimum energy scale of a step but does not by itself prescribe a precision schedule. Fourth, the $B$ parallel trajectories share $z_0$ and diverge gradually, so early rollout steps overlap. Fifth, under sustained back-to-back planning, GPU-class power density raises a mm-scale tile by $256.07\text{ K}$ even with an idealized heat spreader, whereas a $40\times$ lower power density would limit the rise to $6.52\text{ K}$ (Section \ref{subsec:TransientNonLinear}). 

\begin{figure*}[htbp]
\centerline{\fbox{\includegraphics[width=0.8\linewidth]{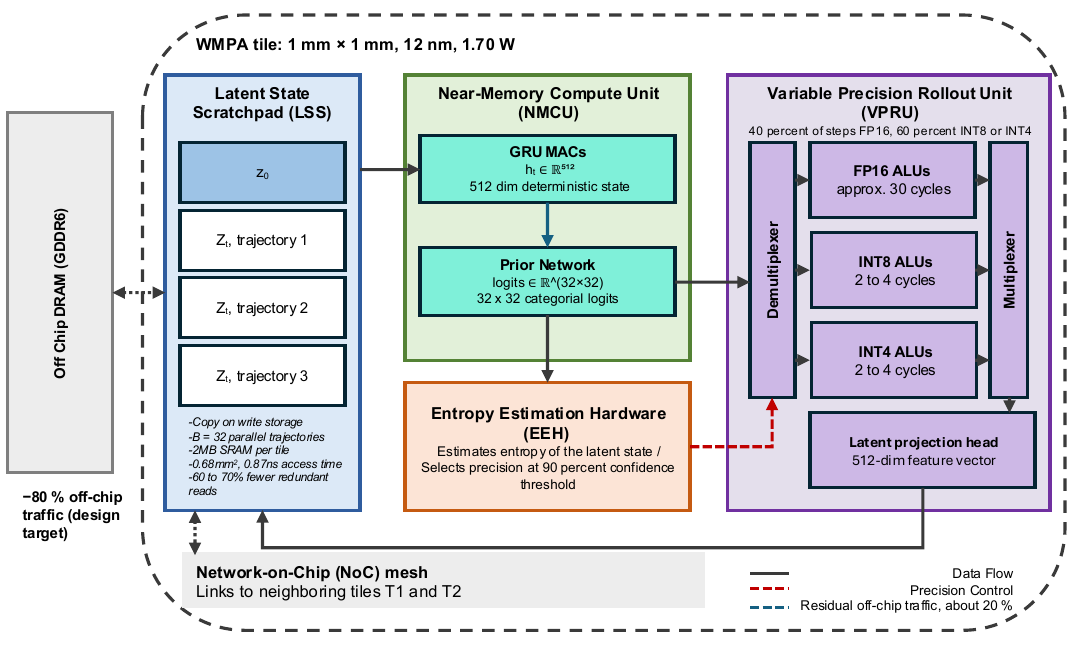}}}
\caption{WMPA block diagram showing the NMCU, EEH, VPRU, and LSS components and the data flow during a rollout step (1.70 W denotes the 40$\times$-reduced package power target, corresponding to 0.154 W per $1\text{ mm}^2$ tile under the hotspot scaling of Section \ref{subsec:TransientNonLinear}).}
\label{fig5}
\end{figure*}

These observations call for an architecture that breaks the bandwidth bottleneck, spends precision where the prediction is uncertain, and fits a thermal budget. The World Model Planning Accelerator (WMPA) is a conceptual co-designed chip with four specialized components, shown in Fig. \ref{fig5}.

\subsection{Component Design and Justification}
\label{subsec:CompDesign}

\textbf{Near-Memory Compute Units (NMCU).} The roofline analysis of Section \ref{subsec:GenRoof} identifies off-chip data movement, not arithmetic, as the bottleneck. The NMCU co-locates compute logic with SRAM, so the GRU state update and latent projection execute inside the memory subsystem without transfers over the main memory bus \cite{mutlu2019processing}. The 80\% cut in off-chip traffic is a design target and not a figure from the literature, and it presumes that the NMCU absorbs the memory-bound GRU and projection operations.

\textbf{Entropy Estimation Hardware (EEH).} The EEH estimates the Shannon entropy of the predictive distribution over the next latent state in real time and sets the numerical precision per step. High-entropy predictions run at full precision and low-entropy (confident) predictions at reduced precision, since a confident prediction is expected to be less sensitive to quantization error. This is a robustness heuristic and not a consequence of the bounds of Section \ref{Sec:TheAnalysis}, which fix only the minimum energy scale. Low-entropy steps are expected to make up a substantial share of a converged rollout (Section \ref{subsec:DynActSparsity}), but the accuracy cost of reduced precision remains to be validated.

\textbf{Variable Precision Rollout Units (VPRU).} Numerical precision trades off against hardware cost. For this small sequential GRU workload we assume that an FP16 tensor-core operation takes about 30 cycles end to end, whereas INT8 ALUs take 2 to 4 cycles at roughly $2\times$ the FP16 throughput and INT4 ALUs deliver roughly $4\times$. Under the $K(b) \propto b^2$ gate-count model (Section \ref{subsec:MACBound}), INT8 and FP16 ALUs cost about $4\times$ and up to $16\times$ the energy per operation of INT4, the latter being an upper estimate. Driven by the EEH, the VPRU uses this trade-off at runtime. As Fig. \ref{fig2} shows, computation goes to the FP16 ALUs while the model's prediction confidence is below a 90\% threshold, as in early exploratory steps, and to the INT8 or INT4 ALUs above it. This progressive precision reduction \cite{micikevicius2018mixed} confines the high cost of full precision to high-entropy states. In the $H = 20$ stress test the confidence crosses the threshold between steps 7 and 8, so 8 of the 20 steps (40\%) run on FP16 ALUs and 12 (60\%) on INT8 or INT4 ALUs. This split comes from a synthetic proxy and not a trained policy, and it must be validated against the rollout dynamics of a trained agent.

\textbf{Latent State Scratchpad (LSS) \& Inter-Tile Routing.} In tree-based planning the $B$ trajectories share action prefixes before branching. The LSS is an on-chip SRAM buffer that stores their latent states with copy-on-write semantics, so shared prefixes are cached and states are duplicated only at branch points, which reduces redundant read bandwidth by an estimated 60 to 70\%. A larger capacity lowers the DRAM miss rate at the cost of access latency, and CACTI 7.0 \cite{balasubramonian2017cacti}, run at 22 nm and scaled to 12 nm, projects that growing the LSS from 1 MB to 2 MB raises its area from 0.32 mm$^2$ to 0.68 mm$^2$ and its access time from 0.65 ns to 0.87 ns (Section \ref{subsec:ProjPerf} gives the scaling). The WMPA targets 2 MB per tile as a balance between reuse and latency. To limit latency, the WMPA partitions the LSS into smaller SRAM arrays distributed across tiles (e.g., T1 and T2) and links them with a high-bandwidth Network-on-Chip (NoC) mesh.

\subsection{Projected Performance, Area, and Emulation}\label{subsec:ProjPerf}

We model the SRAM arrays of the LSS and NMCU with CACTI 7.0 \cite{balasubramonian2017cacti}, which has no validated device data below 22 nm. We therefore model the 2 MB LSS at 22 nm, which gives a data-array area of $2.297\text{ mm}^2$ and an access time of 1.599 ns, and scale by $(12/22)^2$ for area and $12/22$ for access time to obtain $0.68\text{ mm}^2$ and 0.87 ns at 12 nm. This is a first-order approximation and not validated 12 nm compiler output. The SRAM array takes roughly 68\% of the $1\text{ mm} \times 1\text{ mm}$ tile of the thermal simulation (Section \ref{subsec:TransientNonLinear}), leaving the rest for the NMCU, VPRU, and EEH logic.

Data movement is expected to dominate the energy of this workload (Section \ref{subsec:GenRoof}), and for a memory-bound operation compute utilization is bounded by $\mathrm{AI}/\rho$, which on the T4 ($\rho = 203.5$ FLOPs/byte) ranges from $0.19\%$ for stochastic sampling to $7.6\%$ for the GRU transition.
The VPRU routes 40\% of rollout steps to FP16 and 60\% to INT8 or INT4 under the $K(b) \propto b^2$ model, and the LSS cuts redundant read bandwidth by an estimated 60 to 70\%. Together with the NMCU, the WMPA targets about 80\% less off-chip traffic.
The NMCU and LSS act on the same data-movement term, so their savings overlap and cannot be multiplied. We therefore adopt a 40$\times$ reduction in total package power dissipation, from 68.19 W to 1.70 W (corresponding to 0.154 W on the $1\text{ mm} \times 1\text{ mm}$ active tile under the same 9\% hotspot share of Section \ref{subsec:TransientNonLinear}, which assumes the WMPA package distributes power like the T4, or $q_{v,\text{WMPA}} = 1.54 \times 10^9\text{ W/m}^3$), as a design target and not as a derived or measured result, to be replaced by synthesized power numbers. Under the thermal model of Section \ref{subsec:TransientNonLinear} this per-tile dissipation holds the tile at 306.52 K and, if energy per call scales with power, leaves the WMPA about $1.5\times 10^9$ times above the CMOS floor.
Because the one-dimensional balance $\Delta T \approx Q/h_{eff}$ is linear in power, a reduction of about $5\times$ already brings the baseline rise to roughly 50 K, so the thermal conclusion does not depend on reaching the full target.

The WMPA still operates many orders of magnitude above the theoretical bounds, but the design target yields a workload that dedicated silicon can sustain within an idealized thermal budget, a prerequisite for further optimization.
\section{Conclusion}
This paper characterizes the computational and physical cost of world model planning. Combining information-theoretic bounds, hardware profiling, and 3D transient nonlinear thermal simulation, we find that a GPU executing latent rollouts consumes roughly eleven orders of magnitude more energy than the CMOS dynamic floor, and that under sustained back-to-back planning this inefficiency drives a millimeter-scale tile to an unsustainable temperature even with an idealized heat spreader.

Hardware for world model planning should therefore follow the structure of the computation, which is sequential and, in our synthetic proxy, increasingly sparse and low-entropy, and not that of general-purpose accelerators. The WMPA combines near-memory computing with entropy-driven precision scaling to attack the dominant data-movement and thermal constraints. It remains far above the theoretical floor, but under its power-reduction design target it holds the tile near ambient, which gives a foundation for further architectural and device-level optimization. Three limits qualify these results. The precision schedule rests on a synthetic entropy-collapse proxy, the $40\times$ reduction is a design target and not a synthesis result, and the thermal model places an ideal sink behind the spreader. Addressing them requires a trained agent, a synthesized implementation, and a package-level thermal model. We view the WMPA as a starting point for algorithm-hardware co-design that brings world models to power- and thermally constrained robots.

\bibliographystyle{IEEEtran}
\bibliography{ref.bib}

@article{schrittwieser2020mastering,
  title={Mastering atari, go, chess and shogi by planning with a learned model},
  author={Schrittwieser, Julian and Antonoglou, Ioannis and Hubert, Thomas and Simonyan, Karen and Sifre, Laurent and Schmitt, Simon and Guez, Arthur and Lockhart, Edward and Hassabis, Demis and Graepel, Thore and others},
  journal={Nature},
  volume={588},
  number={7839},
  pages={604--609},
  year={2020},
  publisher={Nature Publishing Group}
}

@inproceedings{hansen2022temporal,
  title={Temporal difference learning for model predictive control},
  author={Hansen, Nicklas and Su, Hao and Wang, Xiaolong},
  booktitle={Proceedings of the 39th International Conference on Machine Learning (ICML)},
  volume={162},
  pages={8387--8406},
  year={2022},
  publisher={PMLR}
}

@inproceedings{jouppi2017in,
  title={In-datacenter performance analysis of a tensor processing unit},
  author={Jouppi, Norman P and Young, Cliff and Patil, Nishant and Patterson, David and Agrawal, Gaurav and Bajwa, Raminder and Bates, Sarah and Bhatia, Suresh and Boden, Nan and Borchers, Al and others},
  booktitle={Proceedings of the 44th Annual International Symposium on Computer Architecture (ISCA)},
  pages={1--12},
  year={2017},
  organization={IEEE}
}

@inproceedings{lin2018architectural,
  title={The architectural implications of autonomous driving: Constraints and acceleration},
  author={Lin, Shih-Chieh and Zhang, Yunqi and Hsu, Chang-Hong and Skach, Matt and Haque, Md E and Tang, Lingjia and Mars, Jason},
  booktitle={Proceedings of the 23rd International Conference on Architectural Support for Programming Languages and Operating Systems (ASPLOS)},
  pages={751--766},
  year={2018},
  organization={IEEE}
}

@article{hafner2025mastering,
  title={Mastering diverse control tasks through world models},
  author={Hafner, Danijar and Pasukonis, Jurgis and Ba, Jimmy and others},
  journal={Nature},
  volume={640},
  pages={647--653},
  year={2025},
  publisher={Nature Publishing Group}
}

@article{williams2009roofline,
  title={Roofline: an insightful visual performance model for multicore architectures},
  author={Williams, Samuel and Waterman, Andrew and Patterson, David},
  journal={Communications of the ACM},
  volume={52},
  number={4},
  pages={65--76},
  year={2009},
  publisher={ACM}
}

@article{landauer1961irreversibility,
  title={Irreversibility and heat generation in the computing process},
  author={Landauer, Rolf},
  journal={IBM journal of research and development},
  volume={5},
  number={3},
  pages={183--191},
  year={1961},
  publisher={IBM}
}

@book{cover2006elements,
  title={Elements of information theory},
  author={Cover, Thomas M and Thomas, Joy A},
  year={2006},
  publisher={John Wiley \& Sons}
}

@article{black1969electromigration,
  title={Electromigration—A brief survey and some recent results},
  author={Black, James R},
  journal={IEEE Transactions on Electron Devices},
  volume={16},
  number={4},
  pages={338--347},
  year={1969},
  publisher={IEEE}
}

@article{mutlu2019processing,
  title={Processing data where it makes sense: Enabling in-memory computation},
  author={Mutlu, Onur and Ghose, Saugata and G{\'o}mez-Luna, Juan and Ausavarungnirun, Rachata},
  journal={Microprocessors and Microsystems},
  volume={67},
  pages={28--41},
  year={2019},
  publisher={Elsevier}
}

@inproceedings{micikevicius2018mixed,
  title={Mixed precision training},
  author={Micikevicius, Paulius and Narang, Sharan and Alben, Jonah and Diamos, Gregory and Elsen, Erich and Garcia, David and Ginsburg, Boris and Houston, Michael and others},
  booktitle={International Conference on Learning Representations},
  year={2018}
}

@book{rabaey2002digital,
  title={Digital Integrated Circuits: A Design Perspective},
  author={Rabaey, Jan M and Chandrakasan, Anantha and Nikolic, Borivoje},
  volume={2},
  year={2002},
  publisher={Prentice Hall Englewood Cliffs}
}

@article{glassbrenner1964thermal,
  title={Thermal conductivity of silicon and germanium from 3 K to the melting point},
  author={Glassbrenner, CJ and Slack, Glen A},
  journal={Physical Review},
  volume={134},
  number={4A},
  pages={A1058},
  year={1964},
  publisher={APS}
}

@book{savage1998models,
  title={Models of Computation: Exploring the Power of Computing},
  author={Savage, John E},
  year={1998},
  publisher={Addison-Wesley Longman Publishing Co., Inc.}
}

@article{hinton2015distilling,
  title={Distilling the knowledge in a neural network},
  author={Hinton, Geoffrey and Vinyals, Oriol and Dean, Jeff},
  journal={arXiv preprint arXiv:1503.02531},
  year={2015}
}

@techreport{balasubramonian2017cacti,
  title={CACTI 7: Designing 3D memory and main memory architectures},
  author={Balasubramonian, Rajeev and Jouppi, Norman P and Muralimanohar, Naveen},
  institution={University of Utah / Hewlett Packard Labs},
  year={2017}
}

@inproceedings{tsmc12nm,
  title={A 12nm FinFET technology featuring 2nd generation FinFET for low power and high performance applications},
  author={Liu, S and others},
  booktitle={2017 Symposium on VLSI Technology},
  pages={T18--T19},
  year={2017},
  organization={IEEE}
}

@techreport{nvidia2022h100,
  title={NVIDIA H100 Tensor Core GPU Architecture},
  author={{NVIDIA Corporation}},
  institution={NVIDIA},
  year={2022}
}

\end{document}